\documentclass{aa}  
\usepackage{aalongtable}
\usepackage{aalongtable,lscape}
\usepackage[figuresright]{rotating}
\usepackage{epsfig}
\usepackage{graphicx}
\usepackage{natbib}
\bibpunct{(}{)}{;}{a}{}{,} 
\usepackage{txfonts}
\def\sun{\hbox{$\odot$}}
\def\deg{^\textrm{\scriptsize o}}
\begin{document}
\title{New detections of H$_2$O masers in planetary nebulae  and post-AGB
  stars using the Robledo-70m antenna}

\author{O. Su\'arez \inst{1}
        \and J.F. G\'omez \inst{2}
        \and O. Morata \inst{1}
                }
\offprints{O. Su\'arez}

\institute{Laboratorio de Astrof\'\i sica Espacial y F\'\i sica 
Fundamental, INTA, Apartado 50727, E-28080 Madrid, Spain
        \and 
Instituto de Astrof\'{\i}sica de Andaluc\'{\i}a, CSIC, Apartado 3004,
E-18080 Granada, Spain}
 
\date{Received <date>/Acepted <date>}

\abstract
   {} 
   {We investigated the possible relationship between the evolutionary
     stage of post-AGB stars and planetary nebulae (PNe) and the
     presence of water masers in their envelopes}
   {We have used NASA's 70-m antenna in Robledo de Chavela (Spain) to
     search for the water maser transition at 22\,235.08 MHz, towards
     a sample of 105 sources with IRAS colour characteristic of
     post-AGB stars and PNe at declination $> -32 \deg$. 83\% of the
     sources in the sample are post-AGB stars, 15\% PNe or PN
     candidates, while only 2\% seem to be HII regions.}
   {We have detected five water masers, of which four are reported for
     the first time: two in PNe (IRAS 17443$-$2949 and IRAS
     18061$-$2505), a ``water fountain'' in a post-AGB star (IRAS
     16552$-$3050), and one in a source previously catalogued as a PN,
     but whose classification is uncertain (IRAS 17580-3111).}
    {The unexpected detections of water masers in two objects among the
     small subset of PNe led us to suggest that the PNe harbouring
     water masers are a special type of massive, rapidly evolving
     PNe.}


\maketitle
%
%
\section{Introduction}

Stars with masses between 1 and 8 M$\sun$ start losing their envelope
during the Asymptotic Giant Branch (AGB) stage. At this stage, their
mass-loss rate can be as high as 10$^{-4}$~M$\sun$~yr$^{-1}$. When the mass
loss stops, the star enters the post-AGB stage 
\citep[which lasts for
between 10$^2$ and 10$^4$ years depending on the mass;][]{blocker95b},
and its effective temperature increases until it 
is high enough to ionise the expelled envelope. At this moment the star
enters the Planetary Nebula (PN) stage \citep{kwokan93}.

The circumstellar envelope of the evolved stars provides optimal
conditions to pump several types of masers. During the AGB stage, and
in O-rich envelopes, SiO, OH and H$_{2}$O masers can be pumped
\citep{nyman98,telintel91arti,engels96}. These masers are stratified
in such a way that SiO masers occupy the innermost part of the
envelope, close to the star, water masers are located between 10 and
100 AU from the central star, and OH masers are further away, at
$\simeq 10^4$ AU \citep{reidmoran81}. The standard scenario suggests
that, when the mass-loss typical of the AGB phase stops, masers
disappear sequentially, starting from the innermost ones
\citep{lewis89}. The timescales for the switching off of masers have
been estimated to be $\simeq 10$, 100, and 1000 years after the AGB
mass-loss stops, for SiO, H$_2$O, and OH masers, respectively
\citep{gomezy90}, i.e., mostly during the post-AGB phase.  Therefore,
water masers were thought not to be present during the PNe stage, and
only a few of them would show OH emission \citep{zijlstra89}. However,
\citet{miranda01} reported the first confirmed association of a water
maser with a PN, in K3-35. Since this detection, only one other PN has
been reported to harbour an H$_{2}$O maser: IRAS~17347$-$3139
\citep{itzi04}.

Several surveys have been made to search for water masers in evolved
stars. Most of them examined to OH/IR and Mira stars (i.e., AGB
stars), which have relatively high detection rates \citep[50\%
in][]{engels96}.  These searches for H$_2$O masers were based on IRAS
colours or fluxes of the target sources
\citep{engels84,zucker87,likkel89,deguchi89}, whereas others searched
for this emission in known OH/IR stars \citep{engels96}. While some
sources with IRAS colours typical of OH/IR and Mira stars are post-AGB
stars, no survey so far had focused on the area of the IRAS
colour-colour diagram which is mostly populated by post-AGB stars and
young PNe, until the very recent one by \citet{deacon06}.  Only a few
other works have purposely carried out water maser studies in phases
later than the AGB.  For example, \citet{engels02} performed a
monitoring of water masers in four post-AGB stars, and \citet{itzi04}
searched for water masers in 26 known PNe.

Systematic and sensitive searches for water masers in a large sample
of sources including post-AGB stars and PNe are important to properly
address whether the presence of water masers in these phases is an
evolutionary effect or it is dominated by the characteristics of the
central sources. Moreover, the water masers found in these sources can
be used as a powerful tool to study these important phases of stellar
evolution with an unsurpassable angular resolution ($\la 1$ mas), by
means of radio interferometric observations.

In this work, we have pursued such a large-scale survey, based on the
sources presented in the spectroscopic atlas of post-AGB stars and PNe
of \citet[][hereafter SGM]{suarez06}. The initial goal of this survey
was to find possible relationships between the evolutionary stage of
the post-AGB stars and the presence of water masers in their
envelopes. However, the atlas of SGM also included some young PNe, and
the detections of water masers in two of these sources
(Sec.~\ref{results}) somewhat changed the initially intended scope. We
also note that the survey for water maser emission presented here is
one of the most sensitive ones ever carried out toward evolved stars.

In Sec.~\ref{selection} we describe our selection criteria for the
target sources. In Sec.~\ref{observations} we describe the
observations performed. Section~\ref{results} contains the results
obtained and a description of the detected sources. In
Sec.~\ref{discussion} we discuss our results and in
Sect.~\ref{conclusions} we present our conclusions.

\section{Source sample selection criteria}
\label{selection}

The sample of objects was selected from the atlas of SGM. This atlas
was chosen because it contains the largest number of post-AGB stars
with spectroscopic classification compiled to date. It also contains
some relatively young PNe.

The SGM atlas is composed of IRAS
sources located in the region of the IRAS colour-colour diagram
populated mainly by post-AGB stars and PNe \citep{vanderveen89}. The
sources in this region are characterised by a range of dust
temperatures in their envelopes between 200\,K and 80\,K, with an
expanding radius between 0.01\,pc and 0.1\,pc.

We have observed all the sources in SGM with $\delta > -32\degr$ that
were classified by these authors as evolved sources, independently of
their evolutionary stage. They divide the evolved sources in five
groups: PN, post-AGB stars, transition sources (that are evolving
between these two categories), peculiar sources (evolved sources with
peculiar characteristics), and objects with no optical counterpart.
The total amounted to 105 targets. The complete list of observed
sources, with their coordinates and classifications according to SGM
are listed on Table~\ref{observadas} (in the online version). In this
table we also provide other classifications found in the literature
for the sources in which no optical counterpart was found by SGM. For
the sources with an optical counterpart, we rely on the
classification given by SGM based on their optical spectra.

\begin{figure}[h!]
 \centering
 \includegraphics[width=9cm]{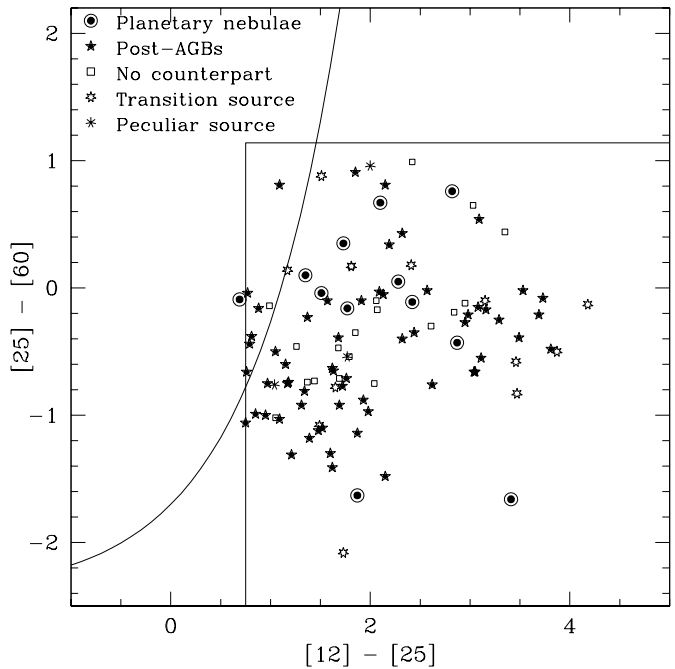}
 \caption{IRAS colour-colour diagram of the observed sources. The
   evolutionary classification follows SGM, and the sources in which
   they did not find an optical counterpart are labelled as such. The
   two straight lines shown in the diagram delimit the selected
   colours and the curve is the line modelled by \citet{bedijn} where
   the AGB stars are located.}
 \label{completo_hn}
\end{figure}

In Fig.~\ref{completo_hn} we represent the observed sources in the
IRAS two-colour diagram showing their evolutionary stage as given by
SGM. The IRAS colours [12]$-$[25] and [25]$-$[60] are defined in the
classical way as:

\begin{equation}
 [12]-[25] = -2.5~ \log \frac{F_{12\rm{\mu}m}}{F_{25\rm{\mu}m}}
\end{equation}

\begin{equation}
[25]-[60] = -2.5~ \log \frac{F_{25\rm{\mu}m}}{F_{60\rm{\mu}m}}.
\end{equation}

The 105 observed targets are initially distributed in evolutionary
classes as follows: 

\begin{itemize}
        \item   61 post-AGB stars with optical counterpart
        \item   13 transition sources
        \item   12 PN with optical counterpart
        \item   19 stars without optical counterpart (2 HII regions,
          13 post-AGB candidates, 4 PN candidates) 
\end{itemize}

This segregation can be further simplified, if we do not use the
information on the presence or not of an optical counterpart.
Moreover, we will consider as post-AGB stars all objects classified by
SGM as ``transition objects'', except for IRAS 17347$-$3139, which
seems to be a young PN \citep{itzi04,gomezpatxi05}. Note that only two
objects in the sample do not seem to be evolved stars.

For the sources without an optical counterpart, we used the
classifications given in the literature, although we note that these
classifications may not be as reliable as for the ones with observed
optical spectra.

With these
considerations and caveats, the percentage of sources in each class is:
\begin{itemize}
\item 81.9~\% post-AGB stars 
\item 16.2~\% PNe (bona-fide and candidates)
\item 1.9~\% HII regions
\end{itemize}

The observed sample included at least three sources previously
confirmed to harbour water masers: the post-AGB stars \object{IRAS
  07331+0021} \citep{engels96} and IRAS 17392$-$3020 \citep{deacon06},
and the PN \object{IRAS 17347-3139} \citep{itzi04}.  \citet{han98}
reported a 4~$\sigma$ detection of 22~Jy toward the post-AGB star IRAS
20406+2953, but \citet{likkel89} had not found any maser emission with
an rms level of 310 mJy and we did not detect it either (rms$\simeq
50$ mJy), so this reported detection remains to be confirmed.

\section{Observations}
\label{observations}

We observed the 6$_{16} \rightarrow$5$_{23}$ transition of the water
molecule (rest frequency 22235.080 MHz), using  NASA's 70-m antenna
(DSS-63) at Robledo de Chavela (Spain), between October 2004 and March
2006.  The 1.3 cm receiver of this 
antenna comprised a cooled high-electron-mobility transistor (HEMT). The
backend used was a 384-channel spectrometer, covering a bandwidth of
16 MHz ($\simeq 216$ km s$^{-1}$ with $\simeq 0.6$ km s$^{-1}$
resolution). At this frequency, the half-power beamwidth of the
telescope is $\simeq 42''$. Spectra were taken in position-switching
mode. Only left circular polarisation was processed.  System
temperatures ranged between 45 and 150 K, depending on elevation and
weather conditions, and the total integration
time was typically 30 min (on+off). The rms pointing accuracy
was better than 10$''$. The data reduction was performed using the
CLASS package, which is part of the GILDAS software.

\section{Results}
\label{results}

\subsection{Survey results}

The results of our survey are summarised in Tables~\ref{detections}
and \ref{nondetections} (Table~\ref{nondetections} in the online
version). We detected water maser emission in five of the 105 observed
sources (see Table~\ref{detections}, and Figs.~\ref{I07331sp} to
\ref{I18061sp}).  Four of them were detected in this survey for the
first time. They are a post-AGB star: \object{IRAS 16552$-$3050}, a
bona-fide PN: \object{IRAS 18061$-$2505}, and two PN candidates
(without optical counterpart): \object{IRAS 17443$-$2949},
\object{IRAS 17580$-$3111}. However, as discussed below, IRAS
17580$-$3111 is not likely to be a PN.

Of the sources in our survey previously known to show water maser
emission, we detected it in the post-AGB star IRAS 07331+0021, but not
in the PN IRAS 17347-3139, nor in the post-AGB stars IRAS
17392$-$3020 and IRAS 20406+2953.

\begin{figure}[h!]
 \centering
 \includegraphics[width=9cm]{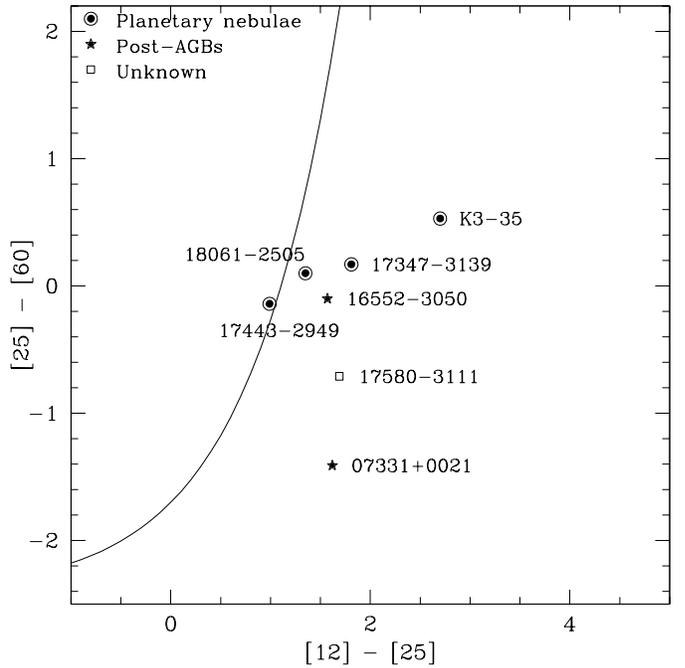}
 \caption{IRAS colour-colour diagram of the sources in which we
   detected H$_2$O maser emission. We also include here the two PNe
   previously known to harbour water masers. The nature of IRAS 17580$-$3111
   is uncertain, although it was previously classified as a PN (see text)}
 \label{completo_hn_detec}
\end{figure}

In
Fig.~\ref{completo_hn_detec} we show their position in the IRAS 
colour-colour diagram, together with their evolutionary stage. We also
include here the position of the other two PNe known to harbour water
masers: K3-35, and IRAS 17347$-$3139. We note that the colours of K3-35 are
contaminated by nearby sources, so its position in this diagram is not
reliable.

Although it would be very interesting to determine whether the
distribution of the detected sources in the IRAS colour-colour diagram
differs from the distribution of the complete sample, the small number
of H$_2$O maser detections precludes us from making an statistically
significant study. The implementations of the 
Kolmogorov-Smirnov test to compare two two-dimensional distributions
require the number of elements in both groups to be $\ga 10$.

\subsection{Detected post-AGB stars}
\subsubsection{\object{IRAS 07331+0021}}

This post-AGB star shows an optical spectrum of K3-5I type (SGM). Its
corresponding low temperature ($\sim$4000~K) indicates that the star
has probably left the AGB phase very recently. This source presents OH
maser emission \citep{telintel91arti} with a standard double-peaked
profile and mean velocity of $\simeq 27.8$ km s$^{-1}$. The presence
of a water maser towards IRAS 07331+0021 was reported by
\citet{engels96}, with a peak flux density of 1.2 Jy (0.9 km s$^{-1}$
resolution) at 29.2 km~s$^{-1}$, observed in October 1990.

\begin{figure}[t!]
\centering
 \includegraphics[width=4.3cm,angle=-90]{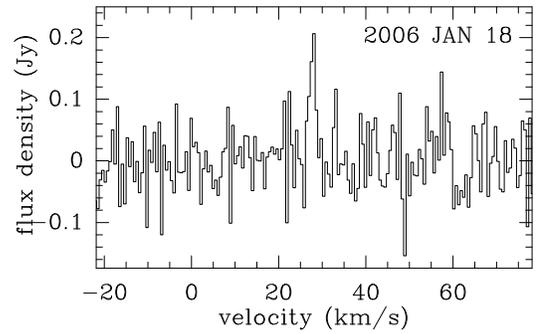}
 \caption{Water maser spectrum of the post-AGB star IRAS 07331+0021}
 \label{I07331sp}
\end{figure}

In our survey, 15 years later, the detected water maser emission was
significantly weaker, with one component of $\simeq 0.21$ Jy at 28.1
km s$^{-1}$ (Fig.~\ref{I07331sp}), close to the mean velocity of the
OH spectrum.

\subsubsection{\object{IRAS 16552$-$3050}}

IRAS 16552$-$3050 was first proposed to be a candidate post-AGB
star by \citet{preite88}, based on its IRAS colours. A possible
optical counterpart of this source was classified by \citet{hu93} as a
K9III type star, and by SGM as a G0I type. However, recent water maser
observations by our group with the VLA (Su\'arez et al., in
preparation) show that there is no optical source at the position of
the water masers.
Previous single-dish observations of the OH maser line at 1612 MHz
\citep{telintel91,hu94} did not show any maser emission, with a
$1\sigma$ rms of 0.2 and 0.1 Jy, respectively.

\begin{figure*}[t!]
\centering
 \includegraphics[width=8cm,angle=-90]{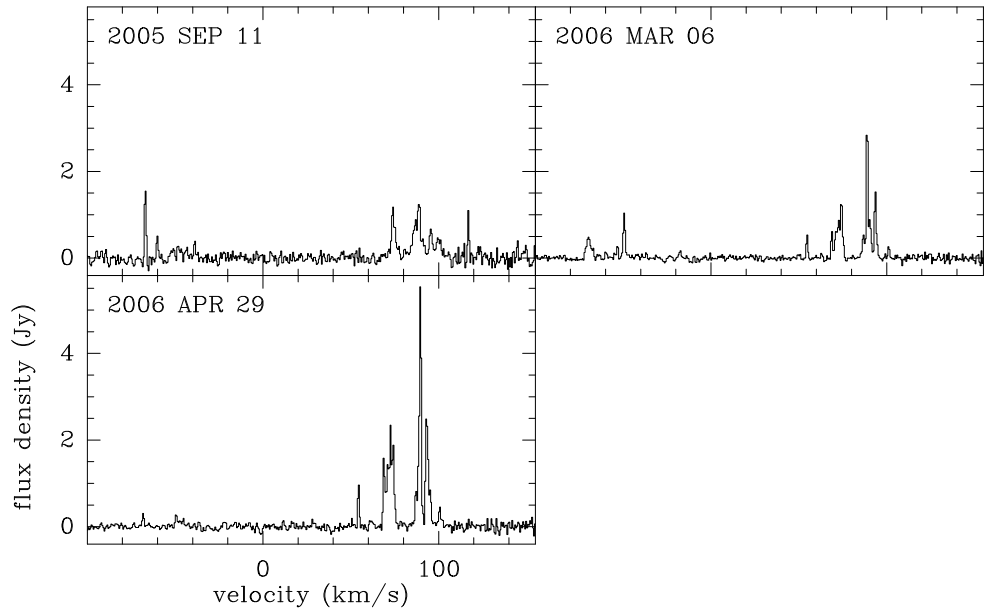}
 \caption{Water maser spectra of the post-AGB star IRAS 16552$-$3050}
 \label{I16552_sp}
\end{figure*}

We detected water maser emission toward IRAS 16552$-$3050 for the
first time on 2005 September 11 (Fig.\ref{I16552_sp}). The spectra
show multiple components, with a maximum velocity separation up to
$\simeq 170$ km s$^{-1}$. These high velocity separations in maser
components indicate that IRAS 16552$-$3050 is the sixth known source
that can be classified as a ``water fountain'' \citep{likkel88}. The
only other known sources of this class are IRAS 16342$-$3814
\citep{sahai99}, IRAS 18043$-$2116 \citep{deacon06} , OH 12.8$-$0.9
\citep{boboltz05}, W43A \citep{imai02}, and IRAS 19134+2134
\citep{imai04}.  Water fountains are evolved sources that have just
started an episode of bipolar mass loss. Since the mass loss that
characterises the Asymptotic Giant Branch (AGB) phase is spherical, in
``water fountains'' we seem to be witnessing the onset of
non-spherical mass loss that may later shape
PNe. 

The spectra taken at different epochs (Fig~\ref{I16552_sp}) show
a high variability in the masers features. In the last epoch
(2006 April 29), the components with negative velocities have nearly
disappeared.

\subsection{Detected PNe and PN candidates}
\subsubsection{\object{IRAS 17443$-$2949}}

This source is one of the objects in the sample of SGM for
which no optical counterpart has been found, and that has been
  classified as a PN.  PNe are usually detected at least in the
strongest nebular lines, such as [OIII] or in H$\alpha$, but in this
case the circumstellar envelope still obscures the radiation from the
central star, even for the nebular lines. Radio continuum emission in
this source was reported by \citet{ratag90} [object RZPM 39], with
$S_\nu({\rm 6 cm})=0.9$ mJy and a diameter of $3\farcs 4$.
\citet{zijlstra89} found OH maser emission in this PN
candidate, and catalogued it as OHPN 10. The distance
  between the radio continuum and the OH positions are $\simeq 9''$,
  which is smaller than the beam of the OH observations
  \citep{zijlstra89}, and therefore, both emissions probably arise
  from the same object.

\begin{figure*}[t!]
\centering
 \includegraphics[width=6cm,angle=-90]{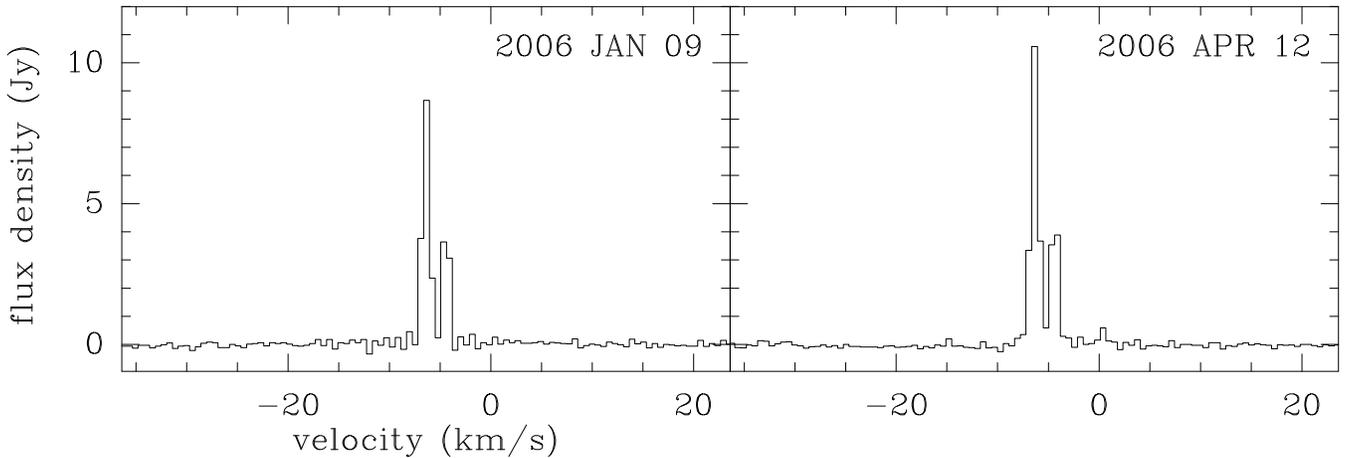}
 \caption{Water maser spectra of the PN IRAS 17443$-$2949}
 \label{I17443sp}
\end{figure*}

We have found water maser emission (Fig.~\ref{I17443sp}) with a
maximum flux density of $\simeq 10$ Jy and two well-defined components
separated by $\simeq 2$ km s$^{-1}$.

\subsubsection{\object{IRAS 18061$-$2505}}

This source is a low-excitation bipolar PN, showing two spectacular
lobes (SGM). The size of the nebula in the optical (H$\alpha$ filter)
is $\simeq 46''\times 15''$. This is the PN with the largest angular
size known to harbour water masers. Its radio continuum emission has
been detected by \citet{condon98} with $S({\rm 18 cm})=3.5 \pm 1.0$
mJy. No OH maser emission was detected in the survey made by
\citet{telintel91arti}, with a $1\sigma$ rms of 0.1~Jy.

\begin{figure*}[t!]
\centering
 \includegraphics[width=8cm,angle=-90]{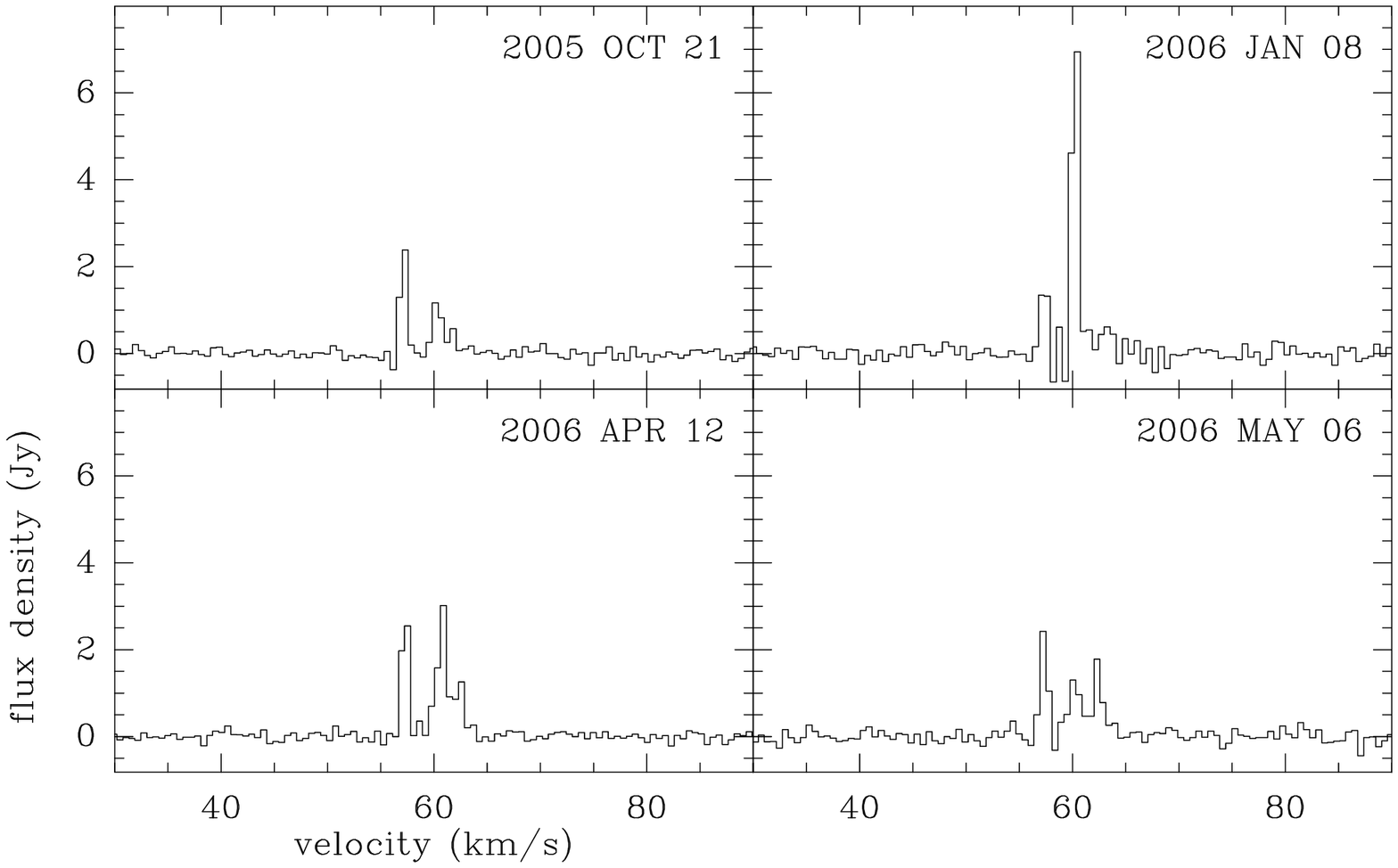}
 \caption{Water maser spectra of the PN IRAS 18061$-$2505}
 \label{I18061sp}
\end{figure*}

The water maser spectra we obtained with the Robledo antenna show
three components (at $\simeq 53.3$, 60.5, and 62.0 km s$^{-1}$) whose
relative intensity is very variable. The peak flux density in the
spectrum was $\simeq 2-3$ Jy in October 2005, April 2006 and May 2006,
but it rose to $\simeq$7~Jy in January 2006 (see
Fig.\ref{I18061sp}).

Recent Very Large Array (VLA) observations have confirmed the spatial
association of the water maser emission with the central star of the PN
\citep*{suarezletter06}.

\subsection{Detected source with uncertain classification: \object{IRAS 17580$-$3111}}

This source was previously classified as a PN, mostly based on
  the presence of radio continuum emission. \citet{ratag90} detected a
  radio continuum source of $S_\nu({\rm 6 cm})=2.5$ mJy, which they
  associated with the IRAS source and catalogued this PN candidate
  as RZPM 45. Although the radio continuum source is within the error
  box of the IRAS position, the more accurate MSX point source catalog
  shows an infrared source (MSX6C G359.7798-04.0728) that is $\simeq
  30''$ away from the radio continuum position, and therefore, their
  association is unlikely (the position of the MSX source has a
  $2\sigma$ error of $\sim 0\farcs 4$). 

OH emission was detected in the region by \citet{zijlstra89},
  who catalogued it as OHPN 11. Because of being single-dish
  observations, it is not possible to determine the location of this
  emission. No optical counterpart has been found by SGM for this
  source.

  In summary, we believe that the identification of IRAS 17580$-$3111
  as a PN is not well justified. A possibility is that this source is
  an OH/IR star. However, we also note that its position in the IRAS
  colour-colour diagram (Fig.  \ref{completo_hn_detec}) is not typical
  of such type of sources, which tend to lie close to the AGB line
  modelled by \citet{bedijn}.  Our speculation is that the source is a
  post-AGB star, given its position in the diagram, although this
  suggestion need confirmation.

\begin{figure}[t!]
\centering
 \includegraphics[width=4.3cm,angle=-90]{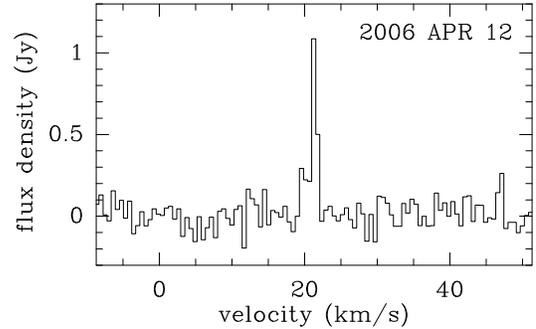}
 \caption{Water maser spectrum of the PN IRAS 17580$-$3111}
 \label{I17580sp}
\end{figure}

We have detected water maser emission (Fig.~\ref{I17580sp}) towards
this source with a flux density of $\simeq 1.1$ Jy. The spectrum is
dominated by a component at $\simeq 21.3$ km s$^{-1}$, although a
weaker one at $\simeq 19.5$ km s$^{-1}$ also seems to be present.
We note that the radio continuum source reported by
  \citet{ratag90} falls outside the Robledo beam, which further
  reinforces our suggestion that the pumping source of the maser
  emission is not likely to be a planetary nebula.

\section{Discussion}
\label{discussion}

\subsection{Characteristics of the detected sources}

The two post-AGB stars in our survey found to harbour water masers do
not seem to have much in common. IRAS 07331+0021 could be a low-mass
star slowly evolving away from the AGB stage. Its position in the IRAS
colour-colour diagram and its low temperature support this hypothesis.
IRAS 16552$-$3050 belongs to the ``water fountain'' type. These
objects evolve rapidly in the post-AGB stage and show a fast stellar
wind that may shape the future PNe.

With our two new detections of water masers in PNe
and PN candidates (hereafter referred to as H$_2$O-PNe),
there are now four such objects known, and we can start
looking for common patterns, although most of these objects have not been
well studied. Two of the four known H$_2$O-PNe
are strongly obscured: IRAS 17443$-$2949 [OHPN 10] has not
been detected in the optical range, while IRAS 17347$-$3139 [OHPN 5]
was only detected in [S~III] at $\lambda$9069~\AA~by SGM, but it does
not show any stellar or nebular emission at shorter wavelengths. The
other two PNe with known water masers are optically visible, although
K3$-$35 is a compact PNe~\citep{miranda98}. IRAS 18061$-$2505 is a
somewhat ``classical'' PN, showing two large bipolar lobes (SGM).

In the three cases for which the PN status has been
  confirmed, and their morphology determined (K3-35, IRAS
17347$-$3139, IRAS 18061$-$2505), the objects are bipolar. Moreover,
all H$_2$O-PNe, except for IRAS 18061$-$2505, show OH maser emission
(see Sect.~\ref{results}).

With so few cases of known H$_2$O-PNe, it is difficult to draw firm
conclusions, although our data seem to suggest that H$_2$O masers
might be favoured in obscured, bipolar PNe, which also show OH maser
emission. The characteristics of these H$_2$O-PNe suggest that they
are very young and relatively massive PNe, as discussed in
\citet{gomezpatxi05} for IRAS 17347$-$3139, which seems to be a good
example of this special type of source. More studies on the physical
properties of these still poorly known PNe in which we have detected
water maser emission would help to confirm the trends that have been
suggested here.

\subsection{Detection rates and completeness of the sample observed}
\label{completeness}

It is surprising that our detection rate of water masers is $\simeq 5$
times larger in objects classified as PNe than in post-AGB stars, and
this seems to contradict the simple scenario of masers disappearing as
stars evolve. The most straightforward conclusion from our detection
rates would be that there is a process that powers water masers in
young PN, and that was not at work in the immediately previous phase.
However, before reaching this or any conclusion based on detection
rates for each category of sources, we first have to analyse the
completeness of the sample observed and any possible biases that may
be present in it.

The sample in SGM aimed to be complete in the selection of post-AGB
stars with optical counterparts. Sources that were not likely to have
optical counterparts were excluded from their catalogue. Since our
original goal in this paper was to study the incidence and evolution
of water masers in post-AGB stars, the SGM catalogue was an
appropriate choice, because it included the largest number of post-AGB
stars with reliable classification. As it was actively avoided by SGM,
the subset of the objects for which no counterpart was found is not
complete.  Regarding PNe, the SGM sample is further biased to favour
faint, low excitation, not well-known PNe. Unfortunately, in two out
of four cases, water masers were found in optically obscured
PNe, exactly the type of source for which the sample is least
complete.

The incompleteness of the samples of PNe and objects without optical
counterpart in our survey, as well as the still uncertain
  nature of the most obscured objects, prevent us from extracting
definitive statistical conclusions from the detection rate results,
however, we can still speculate about possible scenarios for the
presence of water masers in PNe.

\subsection{Position of the detected sources in the IRAS colour-colour diagram}

In Fig.~\ref{completo_hn_detec} we show the position of the detected
sources in the IRAS colour-colour diagram. We also included the
known H$_2$O-PNe IRAS 17347$-$3139 and K3-35, although the latter does
not have reliable colours (see Sect.~\ref{results}). Even though the
location of a star in this diagram does not directly define its
evolutionary status, all the stars located in the same region share
common characteristics. In this case, we do not find any water maser
detection with [12]$-$[25]~$>$~2 or [25]$-$[60]~$>$~0.2, except for
K3-35.

The  region of the diagram in which our detections lie 
is characterised by relatively hot envelopes (or envelopes
close to the central star). The stars located here are those that have
recently left the AGB stage \citep[see][]{bedijn,blocker95b}. If
they already have  high temperatures, this means that they have evolved
very fast, which implies a high mass. This is the case for all the PNe
located in this region.

\subsection{Possible scenarios for the presence of water masers in PNe}

The detection of two water masers towards objects classified as PNe
(thus increasing the number of known H$_2$O-PNe to four sources) is
somewhat surprising if we assume a linear scenario of masers
disappearing after the AGB mass-loss stops \citep{lewis89,gomezy90}.
Even more surprising in this framework could be that the detection
rate in PNe is 5 times higher than in the previous post-AGB phase.
Our results are more easily explained if the water masers that have
been detected in PNe are not the remnants of the masers pumped by the
spherical AGB wind, but they correspond to a later episode of
non-spherical mass loss.

Lewis (1989) pointed out that his evolutionary sequence for
the sequential disappearance of SiO, OH, and H$_2$O masers was valid
for the spherical mass-loss phase only. The discovery of ``water
fountain'' post-AGB stars, in which OH and H$_2$O masers trace
collimated ejections with short dynamical ages is a compelling piece
of evidence of a later phase with non-spherical winds that can pump
water masers \citep[see][]{engels02,deacon06}. Our suggestion is that
water masers found in PNe are related to this phase of non-spherical
winds, and that their precursors in the post-AGB phase could show
water masers of the ``water fountain'' type or similar ones at
somewhat lower velocities.  These suggestions are supported by the
fact that H$_2$O-PN tend to have a bipolar morphology, and that some
of the water maser components in K3-35 have been found at the tips of
the bipolar nebula \citep{miranda01}.

The somewhat higher detection rate we found in PNe of our
sample over that in post-AGB stars can be explained if the pumping of
water masers by non-spherical winds is not a general phase that most
sources undergo, but they pertain to a particular class of evolved
stars. As we pointed out in Sect.~\ref{completeness}, our survey was
not complete for evolved sources without optical counterparts. If the
precursors of H$_2$O-PN are actually post-AGB stars without optical
counterparts, they would have been left out of our observations,
and therefore, the detection rates of water masers in post-AGB
  and PNe in our survey are not directly comparable.

We speculate that these H$_2$O-PNe could represent a
different class of PNe, more massive than the classical ones, whose
evolution through the post-AGB stage is so rapid that they still
maintain their thick AGB envelope when they reach the PN stage. This
envelope would mask the UV radiation from the central star, preventing
the water molecules from being destroyed, while it would also
  tend to obscure their emission in the optical.

If this scenario is true, we should be able to find water maser
emission in the precursors of these obscured and massive PNe. However,
the evolution of these sources would be so rapid during the post-AGB
stage that few objects would be found in this stage. Following
\citet{blocker95b}, a star with 1~M$\sun$ will need
$8.3\times 10 ^3$~yr to reach a temperature of 25\,000~K, while a
star with 5~M$\sun$ will only need 100~yr to reach the same temperature. 
Moreover, these objects would not be visible in the optical,
and their identification as post-AGB stars would be difficult.

As we have already discussed, the sample of observed objects was not
complete in the selection of obscured post-AGB stars. To confirm or
reject our suggestions, a survey for water masers in a complete
set of optically obscured post-AGB stars should be carried out.

\section{Conclusions}
\label{conclusions}

We have performed a survey for water masers in a sample of 105 evolved
stars selected from the SGM atlas, using the Robledo 70-m antenna.

We have detected water masers in five sources, four of which are new
detections. Our new detections comprise one post-AGB ``water
fountain'' source (IRAS 16552$-$3050), two PNe harbouring
  water masers (IRAS 17443$-$2949, and IRAS 18061$-$2505), and a
  source of uncertain nature, previously classified as a PN (IRAS
  17580$-$3111).

   We suggest the possibility that PNe with water maser
     emission are a special class of massive, rapidly evolving PNe
     that are able to maintain a thick envelope, and that water masers
     in these sources are related to processes of non-spherical
     mass-loss, rather than being the remnant of water masers pumped
     by spherical winds in the AGB phase.

\begin{acknowledgements}

  We would like to thank our referee, Dr. Dieter Engels, for his
  careful and useful review, and Dr. Jessica Chapman for her comments
  on the manuscript and for sharing with us some of her results before
  publication.  This paper is based on observations taken during
  ``host-country'' allocated time at Robledo de Chavela.  This time is
  managed by the LAEFF of INTA, under agreement with NASA/INSA. The
  authors would also like to thank the Radio Astronomy Department and
  the operation staff at MDSCC for their support during our
  observations. JFG and OM acknowledge support from MEC (Spain) grant
  AYA 2005-08523-C03 (co-funded by FEDER funds), OS acknowledges
  support from MEC (Spain) grant AYA 2003-09499. OS and JFG are also
  supported by Junta de Andaluc\'{\i}a (grants FQM-1747 and TIC-126).

\end{acknowledgements}

\bibliographystyle{aa}

\clearpage

\clearpage
\onecolumn
{
\begin{longtable}{ccrcc}
\caption{List of observed sources. Classification as in
  \citet{suarez06} is also listed. Other classifications in the
  literature are given for objects listed as peculiar or without an
  optical counterpart in  \citet{suarez06}}  \label{observadas}\\
\hline \hline 
IRAS&RA (J2000)&Dec (J2000)& Classification& Other classification\\
& & & \citet{suarez06}&  \\
\hline
\endfirsthead
\caption[]{List of observed sources (continued).}\\
\hline\hline
IRAS&RA (J2000)&Dec (J2000)& Classification& Other classification\\
& & & [SGM]&  \\
\hline
\endhead
\hline
\endlastfoot

   00509+6623 & 00:54:07.7 &   +66:40:13  & No counterpart &Post-AGB cand.\footnotemark[1] \\        
   01005+7910 & 01:04:45.5 &   +79:26:46  &       Post-AGB &                 \\        
   01259+6823 & 01:29:33.4 &   +68:39:17  &       Post-AGB &                 \\        
   02143+5852 & 02:17:57.8 &   +59:05:52  &       Post-AGB &                 \\        
   04137+7016 & 04:19:07.7 &   +70:23:23  & No counterpart &Post-AGB cand.\footnotemark[2]\\        
& & & & \\                              
   04296+3429 & 04:32:57.0 &   +34:36:12  &       Post-AGB &                 \\        
   05089+0459 & 05:11:36.2 &   +05:03:26  &       Post-AGB &                 \\        
   05113+1347 & 05:14:07.8 &   +13:50:28  &       Post-AGB &                 \\        
   05341+0852 & 05:36:55.1 &   +08:54:09  &       Post-AGB &                 \\        
   05381+1012 & 05:40:57.1 &   +10:14:25  &       Post-AGB &                 \\        
& & & & \\                              
   05471+2351 & 05:50:13.9 &   +23:52:18  &       Peculiar
   &BQ[~]\footnotemark[3]        \\
   05573+3156 & 06:00:33.4 &   +31:56:44  & No counterpart & HII region\footnotemark[4]  \\        
   06499+0145 & 06:52:28.2 &   +01:42:12  & No counterpart & HII region\footnotemark[5] \\        
 06518$-$1041 & 06:54:13.4 & $-$10:45:38  &             PN &                 \\        
 06530$-$0213 & 06:55:31.8 & $-$02:17:28  &       Post-AGB &                 \\        
& & & & \\                              
   06556+1623 & 06:58:30.4 &   +16:19:26  &       Peculiar &BQ[~]\footnotemark[3]        \\
   07134+1005 & 07:16:10.3 &   +09:59:48  &       Post-AGB &                 \\        
 07227$-$1320 & 07:25:03.1 & $-$13:26:20  &       Post-AGB &                 \\        
 07253$-$2001 & 07:27:33.0 & $-$20:07:19  &       Post-AGB &                 \\        
   07331+0021 & 07:35:41.2 &   +00:14:58  &       Post-AGB &                 \\        
& & & & \\                              
   07430+1115 & 07:45:51.4 &   +11:08:19  &       Post-AGB &                 \\        
 08005$-$2356 & 08:02:40.7 & $-$24:04:43  &       Post-AGB &                 \\        
 08187$-$1905 & 08:20:57.1 & $-$19:15:03  &       Post-AGB &                 \\        
 16476$-$1122 & 16:50:24.3 & $-$11:27:58  &       Post-AGB &                 \\        
 16552$-$3050 & 16:58:27.8 & $-$30:55:06  &       Post-AGB &                 \\        
& & & & \\                              
 16559$-$2957 & 16:59:08.2 & $-$30:01:40  & No counterpart & Post-AGB\footnotemark[6]  \\        
 17074$-$1845 & 17:10:24.2 & $-$18:49:01  &Transition      &                 \\        
 17086$-$2403 & 17:11:38.9 & $-$24:07:33  &Transition      &                 \\        
 17106$-$3046 & 17:13:51.8 & $-$30:49:41  &       Post-AGB &                 \\        
 17195$-$2710 & 17:22:43.6 & $-$27:13:37  &       Post-AGB &                 \\        
& & & & \\                              
 17203$-$1534 & 17:23:11.9 & $-$15:37:15  &       Post-AGB &                 \\        
 17223$-$2659 & 17:25:25.2 & $-$27:02:03  &       Post-AGB &                 \\        
 17253$-$2831 & 17:28:33.0 & $-$28:33:26  &       Post-AGB &                 \\        
 17269$-$2235 & 17:29:58.7 & $-$22:37:44  & No counterpart & PN\footnotemark[7]      \\        
 17291$-$2402 & 17:32:12.8 & $-$24:05:00  &Transition      &                 \\        
& & & & \\                              
 17317$-$2743 & 17:34:53.3 & $-$27:45:11  &       Post-AGB &                 \\        
 17332$-$2215 & 17:36:17.7 & $-$22:17:25  &       Post-AGB &                 \\        
 17347$-$3139 & 17:38:00.6 & $-$31:40:55  &Transition      &                 \\        
 17364$-$1238 & 17:39:16.9 & $-$12:40:30  &       Post-AGB &                 \\        
 17371$-$2747 & 17:40:23.3 & $-$27:49:12  &             PN &                 \\        
& & & & \\                              
 17376$-$2040 & 17:40:38.6 & $-$20:41:53  &       Post-AGB &                 \\        
 17381$-$1616 & 17:40:59.8 & $-$16:17:59  &             PN &                 \\        
 17388$-$2203 & 17:41:49.0 & $-$22:05:16  &       Post-AGB &                 \\        
 17392$-$3020 & 17:42:30.4 & $-$30:22:11  &       Post-AGB &                 \\        
 17395$-$0841 & 17:42:14.4 & $-$08:43:19  &             PN &                 \\        
& & & & \\                              
 17423$-$1755 & 17:45:14.2 & $-$17:56:47  &Transition      &                 \\        
   17436+5003 & 17:44:55.5 &   +50:02:39  &       Post-AGB &                 \\        
 17441$-$2411 & 17:47:13.5 & $-$24:12:51  &       Post-AGB &                 \\        
 17443$-$2949 & 17:47:35.3 & $-$29:50:57  & No counterpart & PN\footnotemark[8]        \\        
 17466$-$3031 & 17:49:52.5 & $-$30:33:02  &Transition      &                 \\        
& & & & \\                              
 17488$-$1741 & 17:51:48.9 & $-$17:42:26  &       Post-AGB &                 \\        
 17514$-$1555 & 17:54:21.1 & $-$15:55:52  &             PN &                 \\        
 17516$-$2526 & 17:54:43.4 & $-$25:26:28  &       Post-AGB &                 \\        
 17542$-$0603 & 17:56:56.0 & $-$06:04:10  &       Post-AGB &                 \\        
 17576$-$2653 & 18:00:49.6 & $-$26:53:13  &       Post-AGB &                 \\        
& & & & \\                              
 17579$-$3121 & 18:01:12.8 & $-$31:22:00  &       Post-AGB &                 \\        
 17580$-$3111 & 18:01:19.6 & $-$31:11:22  & No counterpart & PN\footnotemark[8]          \\        
 17582$-$2619 & 18:01:21.6 & $-$26:19:37  & No counterpart & Post-AGB\footnotemark[9]   \\        
 17597$-$1442 & 18:02:38.3 & $-$14:42:03  &             PN &                 \\        
 18044$-$1303 & 18:07:15.3 & $-$13:03:29  &       Post-AGB &                 \\        
& & & & \\                              
 18061$-$2505 & 18:09:12.4 & $-$25:04:34  &             PN &                 \\        
   18062+2410 & 18:08:20.1 &   +24:10:43  &Transition      &                 \\
 18075$-$0924 & 18:10:15.1 & $-$09:23:35  &       Post-AGB &                 \\        
 18186$-$0833 & 18:21:21.1 & $-$08:31:42  &             PN &                 \\        
 18246$-$1032 & 18:27:24.0 & $-$10:30:24  & No counterpart &Non-var OH/IR\footnotemark[9]\\        
& & & & \\                              
 18401$-$1109 & 18:42:57.1 & $-$11:06:53  &             PN &                 \\        
 18415$-$2100 & 18:44:32.0 & $-$20:57:13  &       Peculiar &RCrB type        \\
 18420$-$0512 & 18:44:41.7 & $-$05:09:17  &       Post-AGB &                 \\        
 18442$-$1144 & 18:47:04.0 & $-$11:41:12  &Transition      &                 \\        
   18520+0007 & 18:54:34.8 &   +00:11:04  &             PN &                 \\        
& & & & \\                              
   18524+0544 & 18:54:54.1 &   +05:48:11  & No counterpart & PN\footnotemark[9]          \\        
   18533+0523 & 18:55:46.7 &   +05:27:03  & No counterpart & Post-AGB\footnotemark[9]    \\        
   18576+0341 & 19:00:11.2 &   +03:45:46  & No counterpart & LBV\footnotemark[10]    \\        
   18582+0001 & 19:00:49.0 &   +00:06:14  &       Post-AGB &                 \\        
 19016$-$2330 & 19:04:43.6 & $-$23:26:09  &Transition      &                 \\
& & & & \\                              
   19024+0044 & 19:05:02.1 &   +00:48:51  &       Post-AGB &                 \\        
   19083+0119 & 19:10:54.5 &   +01:24:45  & No counterpart & Post-AGB\footnotemark[6]    \\        
   19114+0002 & 19:13:58.6 &   +00:07:32  &       Post-AGB &                 \\        
   19154+0809 & 19:17:50.6 &   +08:15:08  &             PN &                 \\        
   19200+3457 & 19:21:55.3 &   +35:02:55  &       Post-AGB &                 \\        
& & & & \\                              
   19207+2023 & 19:22:55.8 &   +20:28:55  &       Post-AGB &                 \\        
   19208+1541 & 19:23:05.9 &   +15:47:31  & No counterpart & IR source       \\        
   19225+3013 & 19:24:26.9 &   +30:19:27  &       Post-AGB &                 \\        
   19306+1407 & 19:32:55.1 &   +14:13:37  &       Post-AGB &                 \\        
   19356+0754 & 19:38:02.2 &   +08:01:34  &       Post-AGB &                 \\        
& & & & \\                              
   19386+0155 & 19:41:08.3 &   +02:02:31  &       Post-AGB &                 \\        
   19422+1438 & 19:44:31.7 &   +14:45:25  &       Post-AGB &                 \\        
   19454+2920 & 19:47:24.3 &   +29:28:12  & No counterpart & Post-AGB\footnotemark[6]    \\        
   19477+2401 & 19:49:54.9 &   +24:08:53  &       Post-AGB &                 \\        
 19500$-$1709 & 19:52:52.7 & $-$17:01:50  &       Post-AGB &                 \\        
& & & & \\                              
   19589+4020 & 20:00:43.0 &   +40:29:10  &       Post-AGB &                 \\        
   19589+3419 & 20:00:52.9 &   +34:28:22  &             PN &                 \\        
 19590$-$1249 & 20:01:49.8 & $-$12:41:18  &Transition      &                 \\        
   20160+2734 & 20:18:05.9 &   +27:44:04  &       Post-AGB &                 \\        
   20174+3222 & 20:19:27.8 &   +32:32:15  & No counterpart & Post-AGB\footnotemark[9]   \\        
& & & & \\                              
   20259+4206 & 20:27:42.3 &   +42:16:44  &       Post-AGB &                 \\        
   20406+2953 & 20:42:46.0 &   +30:04:06  & No counterpart & PN cand.\footnotemark[7], post-AGB\footnotemark[9] \\
   20462+3416 & 20:48:16.6 &   +34:27:24  &Transition      &                 \\        
   20572+4919 & 20:58:55.6 &   +49:31:13  &       Post-AGB &                 \\        
   21289+5815 & 21:30:22.8 &   +58:28:52  &       Post-AGB &                 \\        
& & & & \\                              
   21537+6435 & 21:55:04.5 &   +64:49:54  & No counterpart & Post-AGB\footnotemark[9]    \\        
   21546+4721 & 21:56:33.0 &   +47:36:13  &Transition      &                 \\        
   22023+5249 & 22:05:30.3 &   +53:21:33  &Transition      &                 \\        
   22036+5306 & 22:05:30.3 &   +53:21:33  &       Post-AGB &                 \\        
   22223+4327 & 22:24:31.4 &   +43:43:11  &       Post-AGB &                 \\        
\end{longtable}

\noindent
$^1$~\citet{telintel91arti}.\\
$^2$~\citet{vandesteene95}.\\
$^3$~B type star with forbidden emission lines and variability.\\
$^4$~\citet{persi94}.\\
$^5$~\citet{bronfman96}.\\
$^6$~\citet{hu93}.\\
$^7$~\citet{kerber03}.\\
$^8$~\citet{zijlstra89}.\\
$^9$~\citet{gl97}.\\
$^{10}$~Luminous blue variable \citep{ueta01}.\\


}


\clearpage
        \begin{table*}
\begin{center}
\caption{Water maser detections \label{detections}}
\begin{tabular}{ccccccc}
\hline \hline 
IRAS & $V_{\rm min}$\footnotemark[1] & $V_{\rm max}$\footnotemark[2] &
 $V_{\rm peak}$\footnotemark[3] & $S_{\rm
   max}$\footnotemark[4] & $\int S_\nu
 d\nu$\footnotemark[5] &
 Date\footnotemark[6] \\
&  (km~s$^{-1}$) & (km~s$^{-1}$) & (km~s$^{-1}$) & (Jy) & (Jy~km~s$^{-1}$)& \\
\hline

07331+0021  & -104.6 & 111.2 & $28.1\pm 0.6$ &
$0.21\pm 0.10$ & $0.30\pm 0.21$ & 2006-JAN-18\\
16552$-$3050  & -108.2 & 175.2 & $-66.8\pm 0.6$ &
$1.55\pm 0.23$  &  $16.9\pm 1.6$   & 2005-SEP-11\\
           & -215.9 & 215.6 & $88.6\pm 0.6$
& $2.84\pm 0.12$ & $17.8\pm 0.9$ & 2006-MAR-06\\
           & -215.6 & 215.4 & $89.5\pm 0.6$
& $5.54\pm 0.15$ & $28.7\pm 1.5$ & 2006-APR-29\\
17443$-$2949  & -109.1 & 106.6 & $-6.3\pm 0.6$
& $8.67\pm 0.18$ & $13.5\pm 0.6$ & 2006-JAN-09\\
             & -108.0 & 107.7 & $-6.4\pm 0.6$
& $10.58\pm 0.17$ & $15.4\pm 0.5$ & 2006-APR-12\\
17580$-$3111 & -162.1 & 107.7 & $21.3\pm 0.6$ 
& $1.09\pm 0.19$ & $1.3\pm 0.4$ & 2006-APR-12\\
18061$-$2505 & -108.0 & 175.4 & $57.3\pm 0.6$
& $2.4\pm 0.3$ & $3.7\pm 0.7$ & 2005-OCT-21\\
             & -107.7 & 108.1 & $60.5\pm 0.6$
& $6.95\pm 0.24$ & $8.9\pm 0.7$ & 2006-JAN-08\\
             & -110.0 & 105.7 & $60.9\pm 0.6$ 
& $3.01\pm 0.16$ & $7.6\pm 0.4$ & 2006-APR-12\\
             & -105.8 & 109.9 & $57.2\pm 0.6$
& $2.4\pm 0.3$ & $6.074\pm 0.7$ & 2006-MAY-06\\

\hline
\end{tabular}
\end{center}
$^1$ {Minimum velocity covered by the observed
  bandwidth.}\\
$^2$ {Maximum velocity covered by the observed
  bandwidth.}\\
$^3$ {Velocity of the emission peak.}\\
$^4$ {Flux density of the emission peak.}\\
$^5$ {Velocity-integrated flux density of the spectrum.}\\
$^6$ {Date of observation.}\\
\end{table*}


        \begin{longtable}{ccccl}
\caption{Sources without detectable water maser emission \label{nondetections}}\\
\hline \hline 
IRAS &  $V_{\rm min}$\footnotemark[1] & $V_{\rm max}$\footnotemark[2] & rms\footnotemark[3] & Date\footnotemark[4] \\
&  (km~s$^{-1}$) & (km~s$^{-1}$) & (Jy) & \\
\hline
\endfirsthead
\caption[]{Non-detections (continued)}\\
\hline\hline
IRAS &  $V_{\rm min}$ & $V_{\rm max}$ & rms & Date\\
&  (km~s$^{-1}$) & (km~s$^{-1}$) & (Jy) & \\
\hline
\endhead
\hline
\endlastfoot

00509+6623   &  -107.7 & 108.1 & 0.04 & 2005-MAY-20\\
01005+7910   &  -121.5 &  94.3 & 0.09 & 2004-DEC-01\\
             &  -107.7 & 108.0 & 0.08 & 2005-MAY-12\\
             &  -107.7 & 108.0 & 0.06 & 2005-MAY-29\\
01259+6823   &  -107.8 & 107.9 & 0.07 & 2005-FEB-22\\
& & & & \\
02143+5852   &  -107.8 & 107.9 & 0.07 & 2005-FEB-22\\
04137+7016   &  -108.0 & 107.7 & 0.07 & 2004-DEC-01\\
04296+3429   &  -107.8 & 107.9 & 0.06 & 2005-MAR-13\\
05089+0459   &  -107.6 & 108.1 & 0.07 & 2005-MAY-29\\
             &  -147.4 & 107.7 & 0.04 & 2006-FEB-02\\
& & & & \\
05113+1347   &  -108.1 & 107.7 & 0.05 & 2006-JAN-11\\
05341+0852   &  -107.6 & 108.1 & 0.06 & 2005-MAY-29\\
05381+1012   &  -107.9 & 107.8 & 0.04 & 2005-MAR-13\\
05471+2351   &  -108.0 & 107.7 & 0.05 & 2005-FEB-02\\
05573+3156   &  -108.1 & 107.7 & 0.05 & 2005-FEB-02\\
& & & & \\
06499+0145   &  -124.0 &  91.7 & 0.03 & 2005-JAN-03\\
06518$-$1041 &  -108.2 & 107.6 & 0.03 & 2005-JAN-03\\
06530$-$0213 &  -108.0 & 107.7 & 0.04 & 2005-FEB-02\\
06556+1623   &  -108.2 & 107.5 & 0.04 & 2005-FEB-03\\
07134+1005   &  -108.0 & 107.7 & 0.04 & 2005-MAR-12\\
& & & & \\
07227$-$1320 &  -108.1 & 107.7 & 0.04 & 2005-MAR-12\\
07253$-$2001 &  -108.2 & 134.6 & 0.05 & 2005-MAR-12\\
07430+1115   &  -106.0 & 109.8 & 0.03 & 2005-FEB-04\\
             &  -108.1 & 107.6 & 0.04 & 2005-MAR-12\\
08005$-$2356 &  -108.1 & 107.6 & 0.06 & 2005-MAR-12\\
& & & & \\
08187$-$1905 &  -108.1 & 107.6 & 0.05 & 2005-MAR-12\\
16476$-$1122 &  -107.8 & 108.0 & 0.04 & 2006-JAN-08\\
16559$-$2957 &  -107.9 & 107.8 & 0.14 & 2005-SEP-11\\
17074$-$1845 &  -108.1 & 107.6 & 0.10 & 2005-JUL-27\\
17086$-$2403 &  -107.7 & 108.0 & 0.06 & 2006-JAN-08\\
& & & & \\
17106$-$3046 &  -107.7 & 108.0 & 0.10 & 2006-JAN-08\\
17195$-$2710 &  -107.7 & 108.0 & 0.07 & 2006-JAN-08\\
17203$-$1534 &  -108.1 & 107.6 & 0.10 & 2005-JUL-27\\
17223$-$2659 &  -107.7 & 108.0 & 0.12 & 2006-JAN-08\\
             &  -109.3 & 106.4 & 0.05 & 2006-MAR-12\\
& & & & \\
17253$-$2831 &  -107.7 & 108.0 & 0.08 & 2006-JAN-09\\
17269$-$2235 &  -107.7 & 108.0 & 0.13 & 2006-JAN-08\\
             &  -107.9 & 107.8 & 0.04 & 2006-MAR-12\\
17291$-$2402 &  -107.7 & 108.0 & 0.08 & 2006-JAN-08\\
             &  -108.4 & 107.3 & 0.05 & 2006-MAR-12\\
& & & & \\
17317$-$2743 &  -107.7 & 108.0 & 0.08 & 2006-JAN-09\\
17332$-$2215 &  -107.8 & 107.9 & 0.05 & 2006-FEB-21\\
17347$-$3139 &  -107.8 & 107.9 & 0.3  & 2005-OCT-21\\ 
             &  -110.3 & 105.4 & 0.14 & 2006-MAY-06\\
17364$-$1238 &  -107.7 & 108.0 & 0.04 & 2006-JAN-09\\
& & & & \\
17371$-$2747 &  -107.7 & 108.0 & 0.08 & 2006-JAN-09\\
17376$-$2040 &  -107.6 & 108.1 & 0.06 & 2006-FEB-21\\
17381$-$1616 &  -107.9 & 107.8 & 0.03 & 2006-MAR-12\\
17388$-$2203 &  -108.0 & 107.7 & 0.05 & 2006-APR-05\\
17392$-$3020 &  -108.0 & 107.8 & 0.14 & 2005-SEP-11\\
& & & & \\
17395$-$0841 &  -107.9 & 107.8 & 0.04 & 2006-MAR-12\\
17418$-$2713 &  -107.8 & 107.9 & 0.08 & 2006-FEB-21\\  
17423$-$1755 &  -108.0 & 107.7 & 0.05 & 2006-APR-05\\
17436+5003   &  -107.9 & 107.8 & 0.03 & 2005-MAR-13\\
17441$-$2411 &  -108.0 & 107.7 & 0.08 & 2006-APR-12\\ 
& & & & \\
17456$-$2037 &  -108.0 & 107.7 & 0.05 & 2006-APR-12\\
17466$-$3031 &  -108.0 & 107.7 & 0.14 & 2006-APR-06\\
17488$-$1741 &  -107.8 & 107.9 & 0.08 & 2005-OCT-15\\
17514$-$1555 &  -107.4 & 108.3 & 0.07 & 2005-OCT-15\\
17516$-$2526 &  -107.8 & 107.9 & 0.10 & 2005-OCT-21\\
& & & & \\
17542$-$0603 &  -108.0 & 107.7 & 0.03 & 2006-APR-05\\
17576$-$2653 &  -107.7 & 108.1 & 0.09 & 2006-JAN-09\\
17579$-$3121 &  -108.0 & 107.7 & 0.15 & 2006-APR-06\\
17582$-$2619 &  -108.0 & 107.7 & 0.07 & 2006-APR-12\\
17597$-$1442 &  -108.1 & 107.6 & 0.09 & 2005-JUL-27\\ 
& & & & \\
18044$-$1303 &  -107.2 & 108.6 & 0.10 & 2005-OCT-15\\
18062+2410   &  -107.8 & 107.9 & 0.06 & 2005-MAR-13\\ 
18075$-$0924 &  -108.4 & 107.3 & 0.03 & 2006-MAR-12\\
18186$-$0833 &  -108.8 & 107.0 & 0.03 & 2006-MAR-12\\
18246$-$1032 &  -107.9 & 107.8 & 0.04 & 2006-MAR-29\\
& & & & \\
18276$-$1431 &  -108.1 & 107.6 & 0.09 & 2005-JUL-28\\
18401$-$1109 &  -107.9 & 107.9 & 0.04 & 2006-MAR-29\\
18415$-$2100 &  -108.0 & 107.7 & 0.07 & 2005-SEP-11\\
             &  -107.9 & 107.8 & 0.05 & 2006-APR-12\\
18420$-$0512 &   -84.3 & 131.4 & 0.06 & 2005-OCT-15\\
& & & & \\
18442$-$1144 &  -107.9 & 107.9 & 0.04 & 2006-MAR-29\\
18520+0007   &   -24.1 & 191.7 & 0.06 & 2005-JUN-23\\
             &  -107.8 & 108.0 & 0.03 & 2006-MAR-06\\
18524+0544   &  -108.1 & 107.6 & 0.06 & 2005-JUL-28\\
18533+0523   &  -108.2 & 107.5 & 0.07 & 2005-JUL-28\\
& & & & \\
18576+0341   &  -108.2 & 107.6 & 0.07 & 2005-JUL-28\\
18582+0001   &  -161.6 &  54.2 & 0.11 & 2006-FEB-03\\
19016$-$2330 &  -107.8 & 107.9 & 0.09 & 2006-MAR-24\\
19024+0044   &  -108.1 & 107.6 & 0.08 & 2005-JUN-23\\
19083+0119   &  -107.8 & 107.9 & 0.07 & 2005-OCT-15\\
& & & & \\
19114+0002   &  -107.7 & 108.0 & 0.06 & 2006-MAR-12\\
19122$-$0230 &  -107.8 & 107.9 & 0.09 & 2006-MAR-24\\
19154+0809   &  -105.5 & 110.2 & 0.06 & 2005-OCT-15\\
19200+3457   &  -107.7 & 108.0 & 0.06 & 2005-MAR-12\\
19207+2023   &  -108.0 & 107.7 & 0.08 & 2005-JUL-28\\
& & & & \\
19208+1541   &  -108.0 & 107.7 & 0.07 & 2005-JUL-28\\
19225+3013   &  -107.8 & 107.9 & 0.06 & 2004-OCT-29\\
             &  -108.0 & 107.7 & 0.08 & 2005-JUL-28\\
             &  -108.0 & 107.7 & 0.10 & 2005-AUG-10\\
19306+1407   &  -107.9 & 107.9 & 0.06 & 2005-OCT-21\\
& & & & \\
             &  -107.9 & 107.9 & 0.06 & 2006-APR-23\\
19356+0754   &  -108.0 & 107.8 & 0.04 & 2005-SEP-22\\
19386+0155   &  -107.9 & 107.9 & 0.05 & 2006-APR-23\\
19422+1438   &  -107.6 & 108.1 & 0.06 & 2006-JAN-29\\
19454+2920   &  -103.0 & 112.8 & 0.06 & 2005-AUG-26\\
& & & & \\
19477+2401   &  -107.8 & 108.0 & 0.05 & 2006-APR-06\\
19500$-$1709 &  -107.8 & 108.0 & 0.09 & 2006-MAR-24\\
19589+4020   &  -107.7 & 108.0 & 0.05 & 2005-MAR-12\\
19589+3419   &  -108.0 & 107.8 & 0.08 & 2005-AUG-10\\
19590$-$1249 &  -107.9 & 107.9 & 0.05 & 2006-APR-23\\
& & & & \\
20160+2734   &  -107.9 & 107.8 & 0.04 & 2005-SEP-23\\
20174+3222   &  -108.0 & 107.8 & 0.08 & 2005-JUL-28\\
20259+4206   &  -107.7 & 108.0 & 0.07 & 2005-MAR-12\\
20406+2953   &  -108.0 & 107.7 & 0.05 & 2005-AUG-26\\
20462+3416   &  -107.9 & 107.9 & 0.10 & 2005-JUN-23\\
& & & & \\
20572+4919   &   -90.3 & 125.4 & 0.04 & 2005-OCT-15\\
21289+5815   &  -107.8 & 108.0 & 0.12 & 2005-MAY-29\\
21537+6435   &  -110.0 & 105.8 & 0.024& 2005-FEB-21\\
21546+4721   &  -107.7 & 108.1 & 0.03 & 2005-FEB-21\\
22023+5249   &  -107.8 & 108.6 & 0.017& 2005-FEB-26\\
& & & & \\
22036+5306   &  -107.8 & 108.0 & 0.04 & 2006-FEB-03\\
22223+4327   &  -107.6 & 108.1 & 0.04 & 2005-MAR-12\\

\end{longtable}
\noindent
$^1$ {Minimum velocity covered by the observed
  bandwidth.}\\
$^2$ {Maximum velocity covered by the observed
  bandwidth.}\\
$^3$ {Noise level, 1$\sigma$.}\\
$^4$ {Date of observation.}\\

\end{document}